\documentclass[10pt,
twocolumn,
superscriptaddress,
english,
prl,
showpacs,
floatfix,
aps,
]{revtex4-2}
\usepackage[utf8]{inputenc}
\usepackage{amsmath}
\usepackage{amssymb}
\usepackage{graphicx}
\usepackage{xspace}
\usepackage[
separate-uncertainty  = true,
uncertainty-separator =  {\pm},
exponent-product  =  \cdot,
quotient-mode  =  fraction
]{siunitx}

\makeatletter
\@ifundefined{textcolor}{}
{%
	\definecolor{BLACK}{gray}{0}
	\definecolor{WHITE}{gray}{1}
	\definecolor{RED}{rgb}{1,0,0}
	\definecolor{GREEN}{rgb}{0,1,0}
	\definecolor{BLUE}{rgb}{0,0,1}
	\definecolor{CYAN}{cmyk}{1,0,0,0}
	\definecolor{MAGENTA}{cmyk}{0,1,0,0}
	\definecolor{YELLOW}{cmyk}{0,0,1,0}
}

\usepackage{soul}
\usepackage{braket}
\usepackage[backref=none,
bookmarksnumbered=true,
bookmarks=true,
bookmarksopen=true,
colorlinks=true,
citecolor=blue,
linkcolor=blue,
anchorcolor=green,
urlcolor=blue,unicode=false]{hyperref}

\let\baraccent=\= 
\renewcommand{\=}[1]{\stackrel{#1}{=}} 

\newcommand{\didv}{\ensuremath{\mathrm{d}I/\mathrm{d}V}\xspace}

\newcommand{\twoH}{$2H$-NbSe$_2$}
\newcommand{\nbse}{NbSe$_2$~}

\makeatother

\usepackage{babel}

\begin{document}
	
	\title{Charge-density-wave control by adatom manipulation and its effect on magnetic nanostructures}
	
	\author{Lisa M. R\"{u}tten}
	\affiliation{\mbox{Fachbereich Physik, Freie Universit\"at Berlin, 14195 Berlin, Germany}}
	
	\author{Eva Liebhaber}
	\affiliation{\mbox{Fachbereich Physik, Freie Universit\"at Berlin, 14195 Berlin, Germany}}
	
	\author{Kai Rossnagel}
	\affiliation{\mbox{Institut für Experimentelle und Angewandte Physik, Christian-Albrechts-Universit\"at zu Kiel, 24118 Kiel, Germany}}
	\affiliation{\mbox{Ruprecht Haensel Laboratory, Deutsches Elektronen-Synchrotron DESY, 22607 Hamburg, Germany}}
		
	\author{Katharina J. Franke}
	\affiliation{\mbox{Fachbereich Physik, Freie Universit\"at Berlin, 14195 Berlin, Germany}}
	\date{\today}
	
	\begin{abstract}
		Charge-density waves (CDWs) are correlated states of matter, where the electronic density is modulated periodically as a consequence of electronic and phononic interactions. Often, CDW phases coexist with other correlated states, such as superconductivity, spin-density waves or Mott insulators. Controlling CDW phases may therefore enable the manipulation of the energy landscape of these interacting states. \twoH\ is a prime example of a transition metal dichalcogenide (TMDC) hosting CDW order and superconductivity. The CDW is of incommensurate nature resulting in different CDW-to-lattice alignments at the atomic scale. Here, we use the tip of a scanning tunneling microscope (STM) to position adatoms on the surface and induce reversible switching of the CDW domains. We show that the domain structure critically affects other local interactions, namely the hybridization of Yu-Shiba-Rusinov (YSR) states, which arise from exchange interactions of magnetic Fe atoms with the superconductor. 
		Our results suggest that CDW manipulation could also be used to introduce domain walls in coupled spin chains on superconductors, potentially also affecting  topological superconductivity. 
	\end{abstract}
	
	%
	%
	\maketitle 

	Correlated states of matter are among the most widely studied phenomena in modern solid-state physics. When combined within the same material, their competition is particularly intriguing and gives rise to a rich phase diagram \cite{Morosan2012, Manzeli2017}. The competition demands for a deep understanding of the relevant energy scales and at the same time opens pathways for tuning the balance of the co-existing phases, possibly also driving phase transitions \cite{Paschen2020}. Among the interesting correlated states, superconductivity and charge-density order take particularly prominent roles.
	
	\twoH\ is a prime example of a material that exhibits both a CDW and superconductivity at low temperatures. In this material, superconductivity is of Bardeen-Cooper-Schrieffer type albeit with a highly anisotropic Fermi-surface leading to a complicated superconducting gap structure \cite{Yokoya2001,Rodrigo2004a,Borisenko2009,Noat2010,Noat2015,Sanna2022}. After intense studies, the origin of the CDW has been attributed to strong momentum-dependent electron-phonon interactions and the softening of phonon modes at low temperatures rather than Fermi-surface nesting driving the transition to the charge-density ordered phase \cite{Rossnagel2001,Johannes2006,Arguello2015}. 
	
	Additional interest in the CDW in \twoH\ arises from its incommensurability with the atomic lattice. Different alignments with respect to the lattice have varying energies \cite{Guster2019}, causing the long-range order to fragment into domains separated by topological domain walls \cite{Gye2019}. Defects additionally influence the energy landscape of the CDW alignment. Thus, they may effectively act as pinning centers of the CDW's domains \cite{Arguello2014,Shao2016,Wei2017,Kolekar2018,Oh2020,Gao2021,Wang2023} or even serve as seeds in materials which are close to the phase transition, such as the sister compound 2$H$-NbS$_2$ \cite{Wen2020}. Additionally, a recent study using terahertz pulses coupled into an STM junction showed that defects play a crucial role in the dynamics of CDW excitations \cite{Sheng2024}. Defect engineering is, thus, a highly promising avenue for controlling charge-density order and possibly even building resonators for CDW excitations.
	
	Yet, while defects are a natural choice for manipulation of CDW phases, bulk defect engineering imparts several disadvantages. First, it is not easy to control their location during the growth process, second they may affect other material properties, and third, it is impossible to change their position in the bulk a posteriori. Hence, alternative approaches for manipulating CDWs have been explored. One suggestion entails the application of strain \cite{Soumyanarayanan2013,Wei2017, Gao2018}. However, it is very difficult to control and determine the strain at the atomic scale. In another avenue, it was shown that lateral voltage pulses applied to a thin flake induced changes in the CDW order \cite{Walker2022}. This method is restricted to thin samples equipped with additional gates. Voltage pulses in the junction of a scanning tunneling microscope have also successfully induced changes in CDW phases \cite{Bischoff2017, Luican-Mayer2019, Song2022, Chazarin2024}. However, this process is statistical in nature, thus imposing challenges on the precise control and reversibility of the switching.

			\begin{figure*}\centering	\includegraphics[width=\linewidth]{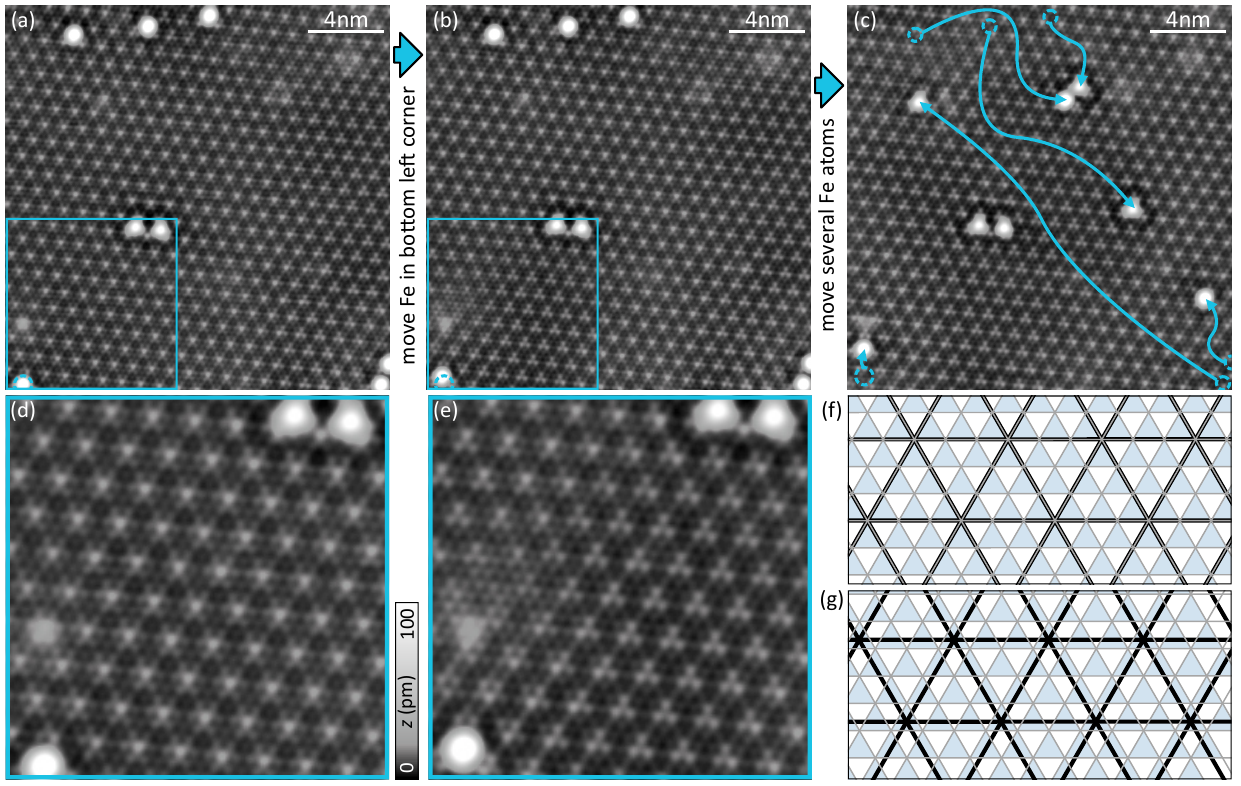}
		\caption{Manipulation of the CDW: (a)-(c) Three stages of manipulating the CDW around a Fe dimer from CC to HC by lateral manipulation of Fe adatoms. In (c) the positions of the Fe atoms prior to manipulation are marked by circles, while the arrow marks their final position. (d) Close-up view on the CC CDW region of the blue square in (a). (e) Close-up of the same region shown in (d) after the CDW was switched to HC in this area. (f, g) Schematic of the CC and HC CDW-to-lattice alignment, respectively, where gray lines indicate the Se lattice, black lines represent the CDW (maxima at crossings) and hollow sites are shaded in blue. Set point: (a)-(e) 10\,mV, 50\,pA. 
					}
		\label{fig:switching}
	\end{figure*}

	Here, we suggest to combine the advantages of defect engineering with the atomic-scale precision of STM. Individual atoms on the surface act as defects that can be positioned at will by dragging them with the STM tip. While we show that adatoms may be used for controlling the domain of structure of the CDW, we take advantage of a second important role of adatoms. When using magnetic atoms on a superconducting CDW material, their exchange-scattering potential causes Yu-Shiba-Rusinov (YSR) states inside the superconducting gap of the substrate \cite{Yu1965,Shiba1968,Rusinov1969,Yazdani1997,Heinrich2018}. Interestingly, the energy of the YSR states and their wave function's symmetry are influenced by the CDW \cite{Liebhaber2020}. The presence of the CDW may thus be a curse and a blessing at the same time: Owing to the variation of the YSR energy with the charge-density modulation, the hybridization properties of closely spaced atoms may be changed. We demonstrate this potential at the example of a Fe dimer adsorbed on \twoH\ upon switching the CDW domain by positioning other adatoms around the dimer.


	STM images of the \twoH\ substrate reveal the atomic structure of the terminating Se layer with an additional apparent-height modulation with a lattice constant of $a_\mathrm{CDW}>3a$ (with $a$ being the Se atomic lattice constant) reflecting the CDW (Fig.\,\ref{fig:switching}a-e). Its incommensurate nature translates into a variation of the alignment of the CDW maxima along the atomic lattice. We show a large-scale topographic image revealing this variation along several domains in the Supplementary Information (SI) (Suppl. Fig.\,1). In case of the CDW maximum coinciding with a Se atom (schematic in Fig.\,\ref{fig:switching}f), we observe a hexagonal pattern as shown in the close up in Fig.\,\ref{fig:switching}d. We refer to this alignment as chalcogen centered (CC). Contrasting this alignment is a hollow-centered (HC) area, which resembles a three-petaled pattern as shown in a close-up in Fig.\,\ref{fig:switching}e and sketched in Fig.\,\ref{fig:switching}g, respectively. The transition between these domains has been assigned to topological domain walls \cite{Gye2019}.  
	
	After deposition at low temperature, we observe two kinds of Fe atoms on the surface (see SI, Note\,2) that differ in apparent height and shape. These types correspond to different adsorptions sites with respect to the crystal lattice: The larger, round adatoms are adsorbed in hollow sites of the Se lattice albeit with a Nb atom underneath. This site has been termed metal site (MS). The smaller, triangular shaped adatoms are located in hollow sites between three Se atoms without an atom below, therefore termed hollow site (HS) \cite{Liebhaber2020, Yang2020}. Both atoms in the center of Fig.\,\ref{fig:switching}a-c are located in hollow sites with respect to the Se lattice. The analysis of the adsorption site is found in the SI, Note 3. They will remain in their precise position in the following and used to probe the effect of a change of the CDW alignment as explained below. 
	
			\begin{figure*}\centering	\includegraphics[width=\linewidth]{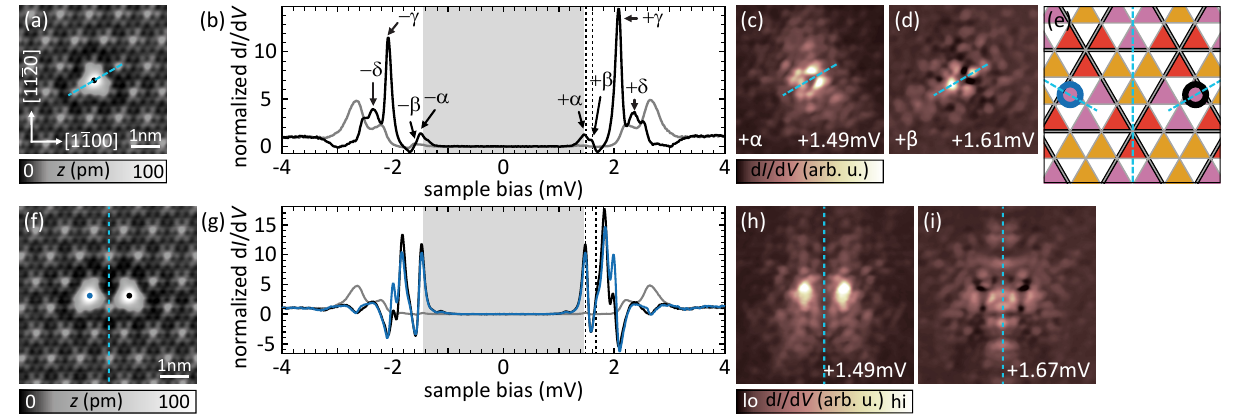}
		\caption{Fe monomer and dimer on the CC CDW domain: (a, f) Topographic images of a Fe atom and a Fe dimer with a spacing of $4a$. All mirror planes are indicated by dashed lines. (b, g) \didv spectra recorded at the positions indicated in (a, f), respectively. Gray traces were recorded on the bare substrate. (c, d, h, i) constant contour \didv maps recorded at the energies of the two energetically lowest resonances of the monomer (c, d) and dimer (h, i). (e) Schematic of the \twoH\ surface in a CC CDW area. Hollow sites are color coded according to their CDW position and the adsorption sites of both atoms are indicated by color-coded circles. All mirror axes present for the individual atoms as well as the dimer are indicated by dashed lines. $\Delta_{\mathrm{tip}}\approx$1.44\,mV; set point: (a), (f) 10\,mV, 50\,pA; rest 5\,mV, 750\,pA; all: V$_\mathrm{rms}$=15\,$\mu$V. 
		}
		\label{fig:before}
	\end{figure*}
		
		To manipulate the CDW alignment, we first move the Fe atom in the bottom left corner closer to the bright defect above it, as indicated by the blue dashed circle that marks the atoms initial position in Fig.\,\ref{fig:switching}a and b. The CDW pattern in the bottom left quadrant (indicated by the blue squares in Fig.\,\ref{fig:switching}a, b and magnified in Fig.\,\ref{fig:switching}d, e) then changes from its hexagonal shape signaling CC alignment (Fig.\,\ref{fig:switching}a, d) to the three-petalled shape of the HC alignment (Fig.\,\ref{fig:switching}b, e). Note, that even the bright defect within this area changes its appearance from hexagonal to triangular. The drastic change caused by a small variation in position of a single Fe atom is rather surprising and indicates energetically-close lying CDW-lattice arrangements. Next, we also move other Fe atoms located close to the edges of our scan frame in Fig.\,\ref{fig:switching}a and b.  A topographic image of the resulting arrangement is presented in Fig.\,\ref{fig:switching}c, where arrows indicate the original positions of all atoms in b. The CDW now assumes the HC configuration in the center of the scan frame (i.e. around the HC Fe dimer). Images of individual manipulation steps are shown in the SI, Note\,7. The switching of the CDW-to-lattice arrangement can be reversed as we show in the SI, Note\,6.

	The possibility to change the CDW domain structure by single atoms is in line with the expectation that defects alter the energy landscape of the CDW alignment. However, we note that not necessarily each change in position of a single atom leads to a rearrangement of the CDW phase. Rather, it is the energy landscape imposed by the distribution of adatoms which dictates the final ground state. Our technique of CDW manipulation thus benefits from a fine energy balance of all atomic potentials in the vicinity. In turn, an assembly of several adatoms may be used to stabilize larger areas of a certain CDW alignment, within which one can freely manipulate single atoms without affecting the CDW. This option is of importance for realizing larger adatom structures with stable properties. Conversely, one may take advantage of the capability to change the CDW and thereby change the properties of the adatom structures. In the following, we show this at the example of the Fe dimer in Fig.\,\ref{fig:switching}. In particular we want to highlight how the CDW switch leads to a change in the properties of hybridized YSR states.

	To fully appreciate the influence of the CDW on coupled YSR states of a Fe dimer, we need to  briefly introduce the YSR states of a single Fe atom and their hybridization characteristics.  Figure\,\ref{fig:before}a shows a topographic image of an isolated Fe atom. This atom is the very same as the right one of the dimer in Fig.\,\ref{fig:switching}a before the left atom was brought into its vicinity. It is adsorbed in a hollow site next to a CDW minimum of the CC alignment. (For a detailed view on the atomic-site determination see SI, Note\,2.) The differential conductance (\didv) spectrum taken on the atom's center is displayed in Fig.\,\ref{fig:before}b (black) together with a spectrum of the bare substrate (gray). The Fe spectrum shows multiple YSR states labeled $\alpha$, $\beta$, $\gamma$, $\delta$ in the sequence of their energy, and in agreement with Ref.\,\cite{Liebhaber2020}. The multiplet is a consequence of four singly-occupied $d$ levels that are crystal-field split and exchange coupled to the substrate \cite{Liebhaber2020, Ruby2016, Choi2017}. Differential conductance maps of the two deepest-lying resonances ($\alpha$ and $\beta$) are shown in Fig.\,\ref{fig:before}c, d and exhibit one mirror plane (dashed line). The reduction from the threefold symmetry of the atomic adsorption site to twofold is caused by the CDW alignment \cite{Liebhaber2020}. The symmetry is visualized in Fig.\,\ref{fig:before}e, where the atomic hollow sites are labeled by colored triangles sitting on the background of the $3\times3$ CDW (thick black lines). The CC CDW-lattice alignment (thick black lines coinciding with thin gray lines of the Se lattice) only preserves one mirror plane for all possible adsorption sites. This symmetry reduction is indicated by a dashed line for the site marked by the black circle, which represents the adsorption site of the Fe atom.

	Figure\,\ref{fig:before}f shows a close up topographic image of the same Fe atom now with an additional Fe atom (indicated by the blue dot) pushed into its vicinity. More precisely, the second atom is located in a hollow site of the Se lattice at a distance of four atomic lattice sites (4$a$) (for structure determination, see SI, Note\,2). We therefore name this arrangement a 4$a$ dimer. It is precisely the one at the center of the scan frame in Fig.\,\ref{fig:switching}a-c. Both atoms of the dimer exhibit similar \didv spectra with an increased number of YSR states as compared to the monomer (Fig.\,\ref{fig:before}g). A doubling of the number of resonances is expected for hybridization. While our energy resolution prohibits a clear identification of eight resonances within the small energy region of the gap, the spatial distribution of the lowest-lying resonances corroborates the interpretation of hybridized YSR states. We show the corresponding \didv\ maps in Fig.\,\ref{fig:before}h and i. The maps feature a nodal plane (Fig.\,\ref{fig:before}h) and maximum intensity (Fig.\,\ref{fig:before}i) perpendicular to the bonding axis. Therefore, we can assign them to the antisymmetric and symmetric linear combination of the monomers YSR wave functions, respectively. A detailed analysis of the adsorption site of the 4$a$ dimer with respect to the CDW reveals that both atoms occupy equivalent sites. The corresponding structure including the CDW is sketched in Fig.\,\ref{fig:before}e with the blue and black circles representing the two atoms. Overall the structure preserves a mirror plane perpendicular to the bonding axis. As a consequence, the hybridized YSR states derived from the symmetric and antisymmetric linear combination of the monomers' states inherit this symmetry \cite{Ruby2018, Liebhaber2022, rutten2024}.

	\begin{figure}\centering	\includegraphics[width=\linewidth]{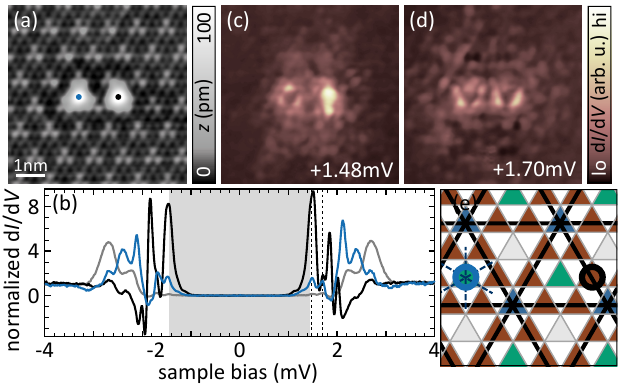}
		\caption{Fe dimer after switching into the HC CDW alignment. (a) Topographic images of the $4a$ dimer after switching the CDW. (b) \didv spectra recorded on both dimer atoms (color coded) and a background trace (gray). (c, d) \didv maps of the two resonances lowest in energy of the $4a$ dimer after the CDW was switched. (e) Schematic of the \twoH\ surface in an HC CDW area. Hollow sites are color coded by their position with respect to the CDW (black lines) and the adsorption sites of the dimer atoms are indicated by color coded circles. All mirror axis present for the individual atoms as well as the dimer are indicated by dashed lines.  $\Delta_{\mathrm{tip}}\approx$1.44\,mV; set point: (a) 10\,mV, 50\,pA; rest 5\,mV, 750\,pA; all: V$_\mathrm{rms}$=15\,$\mu$V.
		}
		\label{fig:after}
	\end{figure}

	Next, we investigate the influence of the change of the CDW from the CC alignment to the HC
	alignment that was induced by manipulation as discussed along Fig.\,\ref{fig:switching}b, c. 
	A close-up view on the dimer after the CDW switch is shown in Fig.\,\ref{fig:after}a with the HC CDW well resolved around the dimer. A detailed analysis of the structure assures that the atoms did not move (see SI Note\,2). The \didv spectra recorded on the atoms have changed dramatically upon the CDW switch (Fig.\,\ref{fig:after}b). While they still exhibit more than four YSR pairs, which is indicative of hybridization, the spectra on the two atoms now differ significantly from one another. The spectrum recorded on the right atom (black trace) shows several intense YSR states deep inside the superconducting gap, while the left one (blue trace) has more intensity close to the coherence peaks in addition to resonances deep inside the gap. 
	
	In agreement with the drastically different spectra, the map of the lowest-lying resonance at  positive bias voltage (Fig.\,\ref{fig:after}c) reflects the loss of mirror symmetry by revealing very different YSR intensities and shapes. The map of the second resonance (Fig.\,\ref{fig:after}d) exhibits more similar patterns around each atom and appears almost symmetric around the dimer's center, yet, there are very faint features that do not obey mirror symmetry. 
	
	To rationalize the observed spectra and their spatial distribution, we analyze the position of the atoms with respect to the CDW (for details see SI, Note\,2). We find that both atoms no longer reside in equivalent sites. The left atom (blue spectrum) sits in a hollow site in a CDW minimum (sketch in Fig.\,\ref{fig:after}e). At this position, all three mirror axes of the Se lattice and CDW align and the symmetry of the crystal is preserved. The right atom (black spectrum) is located in a hollow site on a line connecting two CDW maxima. At this position all mirror symmetries are broken. 
	As a consequence of the different symmetries of the adsorption sites of the individual atoms, there is no mirror axis present in the dimer. 
	
	Analyzing peak positions in the \didv spectra of the dimer shows that the spectra cannot simply be reproduced by adding the spectra of the isolated atoms in the adsorptions sites corresponding to those in the dimer (albeit with different intensity accounting for their spatial decay). Hence, the increase in the number of peaks must originate from hybridization of the inequivalent atoms. Such hybridization patterns are distinct from those that can be obtained in gas-phase molecules because the underlying lattices adds symmetry properties that can be exploited for wave-function design \cite{rutten2024}. To generalize this response to CDW switching, we also investigated a 2$a$ dimer. The observations can be also explained by the different CDW alignments before and after a purposeful switch of the CDW. The results are shown in SI Note\,4.
	
	In conclusion, controlling the position of adatoms on a CDW material is a viable route for manipulating CDW properties. This method can be fine-tuned to obtain energy landscapes that either stabilize larger domains or are on the edge of unstable behavior, allowing for reversible switching. We highlight the potential of the effect of CDW manipulation by showing the drastic changes in YSR hybridization in Fe dimers on \twoH. In combination with previous experiments, where CDW-induced bending of YSR bands was observed \cite{Liebhaber2022}, our findings suggest that CDW manipulation may be used to introduce domain walls in YSR chains or lattices, possibly also affecting the magnetic structure. 
	Furthermore, with the immense interest in topological states of adatom structures \cite{NadjPerge2014, Kim2018, Soldini2023, Yazdani2023}, one may envision to control the topological properties as a consequence of e.g. shifting the band alignment through the Fermi level \cite{Pientka2015, Schneider2021b}.

	\section{Methods}
	
	Our experiments are carried out in a Joule-Thomson STM by Specs, working at 1.1\,K under ultra-high vacuum conditions. We obtain a clean and flat surface by carbon-tape cleaving and deposit Fe atoms directly into the STM at temperatures below 10\,K. To increase the energy resolution of our experiment beyond the Fermi-Dirac limit we pick up a niobium (Nb) crystallite with a W tip, thereby, making our tip apex superconducting. Besides the gain in energy resolution the use of a superconducting tip shifts all spectral features by the superconducting gap of the tip. This superconducting tip gap is indicated by the gray area in all spectra. The Nb also has ideal properties to facilitate lateral manipulation of the Fe atoms. 
	
	\section{Acknowledgements}
	
	We acknowledge financial support by the Deutsche Forschungsgemeinschaft (DFG, German Research Foundation) through Projects No. 277101999 (CRC 183, Project No. C03
) and No. 328545488 (CRC 227, Project No. B05). Sample growth was supported by DFG through Project No. 434434223 (CRC 1461). LMR acknowledges membership by the International Max Planck Research School ``Elementary Processes in Physical Chemistry". 

	%

\clearpage

\setcounter{figure}{0}
\setcounter{section}{0}
\setcounter{equation}{0}
\setcounter{table}{0}
\renewcommand{\theequation}{S\arabic{equation}}
\renewcommand{\thefigure}{S\arabic{figure}}
	\renewcommand{\thetable}{S\arabic{table}}%
	\setcounter{section}{0}
	\renewcommand{\thesection}{S\arabic{section}}%

\onecolumngrid

\maketitle 
\section*{Supplementary Information}
\section{Charge-density-wave control by adatom manipulation and its effect on magnetic
nanostructures}
\section{Supplementary Note 1: Charge-density-wave landscape}

\begin{figure} 
\centering	\includegraphics[width=0.95\linewidth]{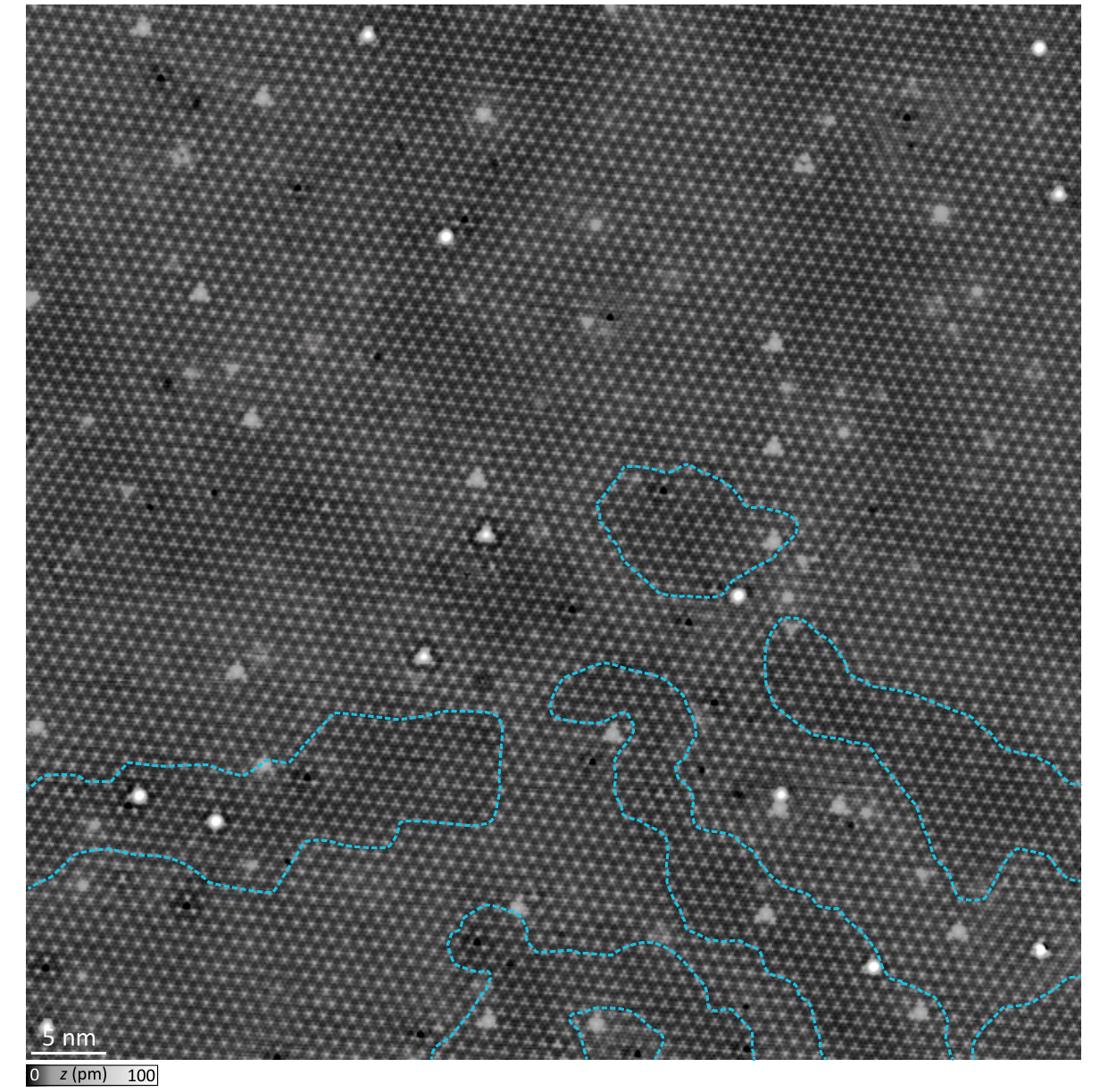}
	\caption{Overview STM image: Large-scale STM image of \twoH\ with atomic resolution after the deposition of Fe atoms. The domains of HC and CC CDW can be seen. As a guide to the eye, we surround HC CDW regions by dashed blue lines in the bottom part of the image.
	 Set point: 10\,mV, 100\,pA.
	}
	\label{fig:ltopo}
\end{figure}

Cleavage of \twoH\ leads to large terraces that can be imaged with atomic resolution using STM. 
Supplementary Fig.\,\ref{fig:ltopo} shows a 70\,nm$\times$70\,nm image of the \twoH\ surface after the deposition of single Fe atoms. The Fe atoms appear as bright protrusions with apparent height $>$100\,pm, which will be discussed in Supplementary Note\,2. 

Notably on this large scale, we observe brighter and darker regions on the surface. We encircle the darker regions by blue dashed lines in the bottom part of the image. We associate the bright and dark regions to different domains of the charge-density wave (CDW) as we can deduce from the atomic resolution. The brighter areas correspond to domains, where the CDW maxima coincide with Se atoms of the terminating layer. Thus, these are called chalcogen-centered (CC) CDW domains as described in more detail in the main text. The darker regions correspond to a hollow-centered (HC) CDW alignment. The overall brighter perception of CC areas can be explained by the larger average apparent height when the CDW and Se atoms are in registry. 

The HC CDW regions appear to have arbitrary shape and size, while all CC CDW regions are connected in agreement with Ref.\,\cite{SGye2019}. If the transitions between CC and HC CDW alignment were purely governed by long-range incommensurability of the CDW, we would expect a periodic landscape of both alignments. The image thus confirms the relevance of local potentials associated to intrinsic defects and adatoms for the CDW-to-lattice alignment.

\section{Supplementary Note 2: As-deposited \NoCaseChange{Fe} atoms}

\begin{figure}\centering	\includegraphics[width=\linewidth]{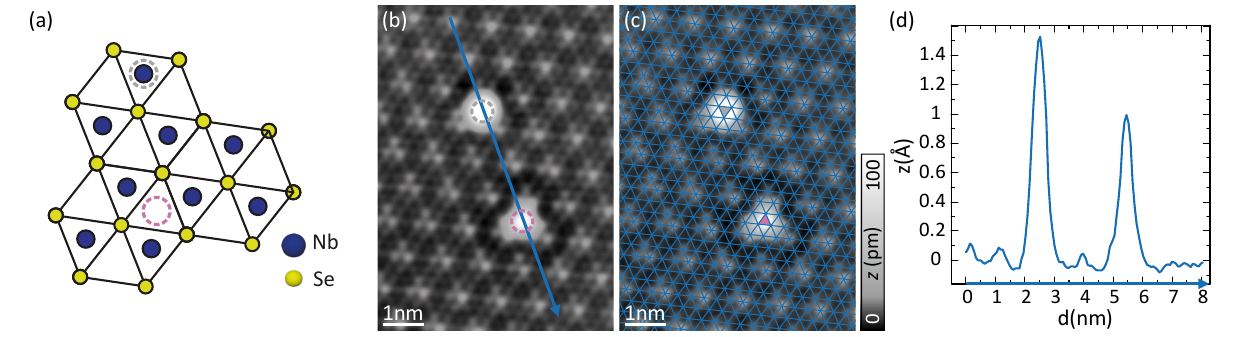}
	\caption{Types of as-deposited Fe atoms: (a) Schematic top view of a \twoH\ surface with the two possible adsorption sites found in experiment indicated by color-coded circles. (b) Topographic image of two Fe atoms of different types. (c) Same image as (b) with an overlayed atomic grid of the Se atoms in the terminating layer. The associated adsorption sites are highlighted by color-coded triangles. (d) Line profile along the arrow in (b), reflecting the distinct apparent heights.
		Set point: (b, c)  10\,mV, 50\,pA.
	}
	\label{fig:sites}
\end{figure}

The as-deposited Fe atoms appear as two different types, which can be distinguished by their apparent height. Supplementary Fig.\,\ref{fig:sites}b shows two atoms with the upper one (marked by a gray circle) appearing as nearly-round protrusion with an apparent height of approximately 150\,pm (line profile along the arrow in Suppl. Fig.\,\ref{fig:sites}d) while the bottom one (marked by a pink circle) appears more triangular and has an apparent height of approximately 110\,pm. 

To determine the adsorption sites of both atoms, we overlay a grid representing the Se lattice to the topographic image (Suppl. Fig.\,\ref{fig:sites}c). We find that both atoms are adsorbed in hollow sites with respect to this lattice. However, the surrounding Se atoms form a triangle pointing downwards around the upper atom and a triangle pointing upwards around the lower one. Referring to a schematic top view of the \twoH\ surface as shown in Suppl. Fig.\,\ref{fig:sites}a we can identify one of the sites as a metal site (Nb atom underneath) and the other one as a hollow site. We can determine which of the sites is a metal and which is a hollow site by looking at the CDW on the left side of the image in Suppl. Fig.\,\ref{fig:sites}b, where the CDW assumes HC alignment. Here, the CDW maximum has to coincide with a hollow site (CDW maxima at metal sites are energetically unfavorable \cite{SGye2019}). As the Se triangles around the CDW maxima are pointing upwards, we can conclude that sites surrounded by upward pointing Se triangles correspond to hollow sites and those surrounded by downward pointing ones correspond to metal sites. As indicated by the color-coded circles in Suppl. Fig.\,\ref{fig:sites}a we can identify the upper (higher) atom as adsorbed in a metal site and the lower (smaller) one as adsorbed in a hollow site in agreement with previous works \cite{SLiebhaber2020, SLiebhaber2022}.

\section{Supplementary Note 3: Determination of adsorption site}

\begin{figure}\centering	\includegraphics[width=\linewidth]{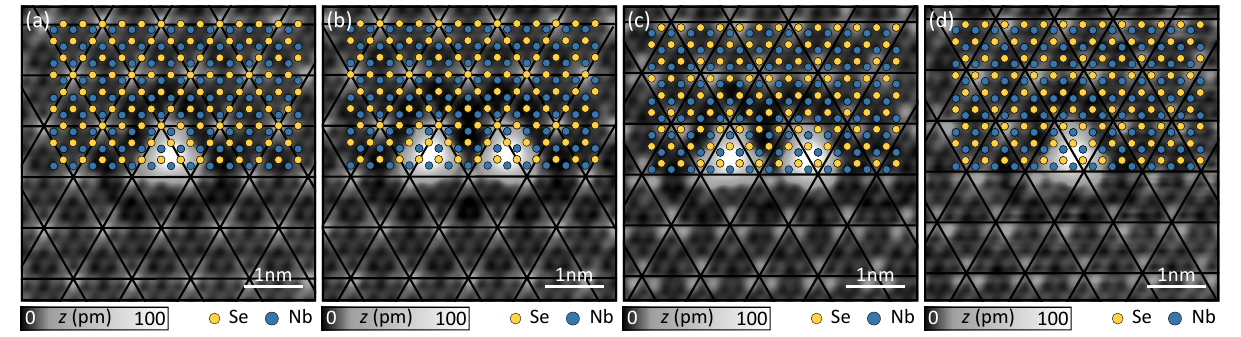}
	\caption{Adsorption-site determination: Topographic images of the monomer (a, d) and the $4a$ dimer (b, c) with the CC (a, b) and the HC (c, d) CDW background. To indicate the adsorption site we superimpose a ball model, where Nb atoms are indicated in blue and Se atoms are indicated in yellow. Additionally, the CDW is indicated by black lines where interceptions correspond to CDW maxima.
		Set point: 10\,mV, 50\,pA.
	}
	\label{fig:lattice}
\end{figure}

We determine the adsorption sites of the Fe atoms discussed in the main text by superimposing atomically-resolved STM images with a ball model for the \nbse\ lattice and a grid of lines for the CDW. Here, yellow and blue balls correspond to selenium and niobium atoms of the topmost atomic layers, respectively, and the interceptions of the grid correspond to CDW maxima. The resulting arrangements for the Fe monomer and the dimer presented in the main text before and after the switching of the CDW are shown in Suppl. Fig.\,\ref{fig:lattice}. Here, Supplementary Fig.\,\ref{fig:lattice}a depicts the single Fe atom with the CC CDW background, (b) depicts the $4a$ dimer with the CC CDW background which has been built by adding a second atom to the left, (c) shows the same $4a$ dimer with the HC CDW background after switching of the CDW, and (d) shows the right atom after the left one has been pushed away by STM manipulation.

The analysis shows that all atoms are adsorbed in hollow sites of the atomic lattice but differ in their position with respect to the CDW. Before switching, i.e., in the CC CDW alignment, both atoms (Suppl. Fig.\,\ref{fig:lattice}a, b) reside in hollow sites next to a CDW minimum (corresponding to atom IV in \cite{SLiebhaber2020}). These sites individually exhibit one mirror axis, as described in detail in the main text. After the switching the CDW, the left atom in (Suppl. Fig.\,\ref{fig:lattice}c) resides in a hollow site at a CDW minimum (same as atom III in \cite{SLiebhaber2020}). In this position all mirror axes of CDW and atomic lattice coincide and the site is threefold symmetric. The right atom sits in a hollow site between two CDW maxima (same as atom VI in \cite{SLiebhaber2020}). In this site, no mirror axis of CDW and atomic lattice coincide, leaving the atom in an asymmetric position.  

After the investigation of the influence of the CDW switch on the dimer properties, we remove the left atom by STM manipulation. The right atom then remains isolated (Suppl. Fig.\,\ref{fig:lattice}d). We note that the CDW was slightly affected by the removal of the left atom. The CDW now slips throughout the image: We find a CC CDW-to-lattice alignment in the top left half, while HC alignment is preserved in the bottom right. This change once more emphasizes the opportunity as well as limitations in the control of CDW manipulation.

\section{Supplementary Note 4: Spectral characteristics of \NoCaseChange{Fe} monomer after switching the CDW}

\begin{figure}\centering	\includegraphics[width=\linewidth]{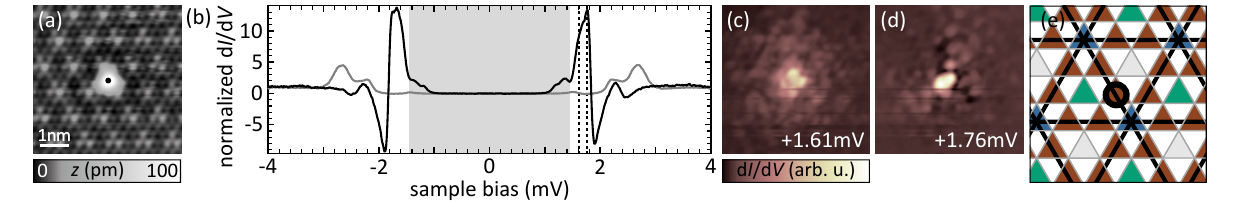}
	\caption{Spectroscopic signature of Fe monomer after CDW switch: (a) Topographic image of the Fe monomer after the CDW was switched from CC to HC. (b) \didv spectra recorded at the position indicated in (a). The gray trace was recorded on the bare substrate. (c, d) Constant-contour \didv maps recorded at the energies of the two energetically lowest resonances. (e) Schematic of the \nbse\ surface in a HC CDW area. Hollow sites are color coded according to their CDW position and the adsorption sites of the atom is indicated by a color coded circle. In this position, all symmetries are broken. 
		$\Delta_{\mathrm{tip}}\approx$1.44\,mV; set point: (a) 10\,mV, 50\,pA; rest 5\,mV, 750\,pA; all: V$_\mathrm{rms}$=15\,$\mu$V.
	}
	\label{fig:monHC}
\end{figure}

We also investigate the spectroscopic signature of the monomer in Suppl. Fig.\,\ref{fig:lattice}d, i.e., after switching the CDW from CC to HC. However, as noted above, investigation of the isolated atom required removal of the left atom of the 4$a$ dimer. This manipulation, unfortunately, led to a slight change of the CDW arrangement compared to the dimer as described above. The \didv spectrum shows broad resonances inside the gap (Suppl. Fig.\,\ref{fig:monHC}b). Due to the limited energy resolution, we cannot identify the individual peaks originating from the crystal-field split $d$ levels. \didv maps on the energy range of the most pronounced broad peak show distinct features (Suppl. Fig.\,\ref{fig:monHC}c, d), suggesting the different $d$ level contributions. The maps do not show any mirror symmetry in accordance with the adsorption site sketched in Suppl. Fig.\,\ref{fig:monHC}e.

\section{Supplementary Note 5: Influence of the CDW switch on a $2a$ dimer}

In the main text, we demonstrate the effect of switching the CDW domain structure on a 4$a$ dimer. Here, we investigate a dimer that consists of two atoms that are only two lattice spacings ($2a$) apart. We note that the area is the same as the one shown in the main text and also the two atoms are the same as in the 4$a$ dimer. In fact, the two dimers have been created by manipulating the two atoms (each marked by a blue and black circle for its identity, respectively) into different sites. After we recorded the data set on the $2a$ dimer we moved the atom marked by the blue circle and formed the $4a$ dimer discussed in the main text. We then switched the CDW around the $4a$ dimer and finally removed the atom marked by the blue dot to study the monomer. We then rebuilt the $2a$ dimer to be able to compare the 2$a$ dimer in the two CDW domains. 

\begin{figure}\centering	\includegraphics[width=\linewidth]{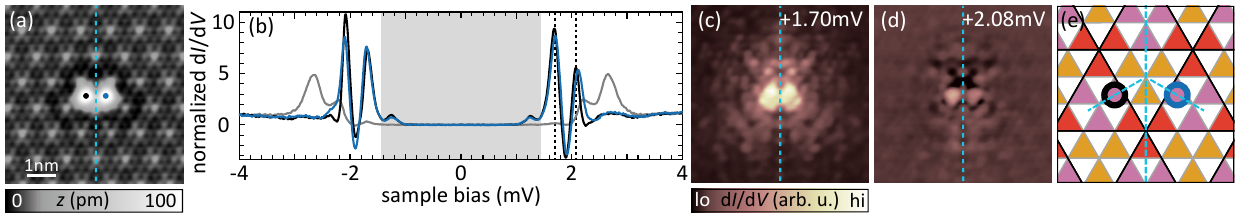}
	\caption{2$a$ dimer in CC CDW domain: (a) Topographic image of a Fe dimer with a spacing of $2a$. (b, g) \didv spectra recorded at the positions indicated in (a). Gray traces were recorded on the bare substrate. (c, d) Constant contour \didv maps recorded at the energies of the two energetically lowest resonances of the dimer. (e) Schematic of the \nbse\ surface in a CC CDW area. Hollow sites are color coded according to their CDW position and the adsorption sites of both atoms are indicated by color coded circles.
		$\Delta_{\mathrm{tip}}\approx$1.44\,mV; Set point: (a) 10\,mV, 50\,pA; rest 5\,mV, 750\,pA; all: V$_\mathrm{rms}$=15\,$\mu$V
	}
	\label{fig:CCbefore}
\end{figure}

The STM image of the 2$a$ dimer in the CC domain is shown in Suppl. Fig.\,\ref{fig:CCbefore}a. Analysis of the precise adsorption sites leads to the structure sketched in Suppl. Fig.\,\ref{fig:CCbefore}e: Both atoms are located in hollow sites close to a CDW minimum. Each site has one mirror plane as indicated by the dashed lines. Additionally, there is a mirror plane between the two atoms. 
\didv spectra on the two atoms are shown in Suppl. Fig.\,\ref{fig:CCbefore}b. The resonances are distinct from the ones seen on the single Fe atom, suggesting some interaction. However, the broad resonances prohibit a detailed analysis of the YSR hybridization. \didv maps recorded at two pronounced peaks reveal the expected mirror symmetry.

\begin{figure}\centering	\includegraphics[width=\linewidth]{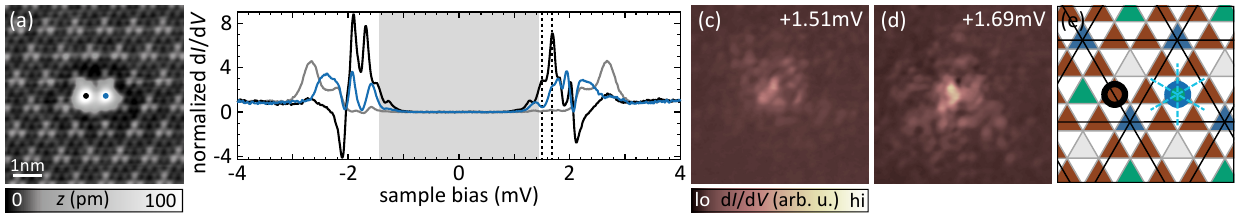}
	\caption{2$a$ dimer in HC CDW domain: (a) Topographic image of a Fe dimer with a spacing of $2a$. (b, g) \didv spectra recorded at the positions indicated in (a). Gray traces were recorded on the bare substrate. (c, d) Constant contour \didv maps recorded at the energies of the two energetically lowest resonances of the dimer. (e) Schematic of the \nbse~surface in a HC CDW area. Hollow sites are color coded according to their CDW position and the adsorption sites of both atoms are indicated by color coded circles.
		$\Delta_\mathrm{tip}\approx$1.44\,mV; set point: (a, f) 10\,mV, 50\,pA; rest 5\,mV, 750\,pA; all: V$_\mathrm{rms}$=15\,$\mu$V.
	}
	\label{fig:HC2a}
\end{figure}

After switching the CDW to the HC alignment, the 2$a$ dimer was re-assembled (after the investigation of the 4$a$ dimer) and is shown in Fig.\,\ref{fig:HC2a}. The \didv spectra differ on the two atoms in agreement with the maps showing unequal intensity around the two atoms. The absence of any mirror symmetry in the dimer is a consequence of the adsorption sites (Fig.\,\ref{fig:HC2a}e).

\section{Supplementary Note 6: Reversibility of CDW switching}

In order to demonstrate the reversibility of the CDW switch, we once again consider the surrounding of the $2a$ dimer as shown in Suppl. Fig.\,\ref{fig:switching}a. The central region of the scan frame (around the dimer) exhibits HC CDW-to-lattice alignment and is surrounded by CC CDW regions. We switch the CDW back to CC alignment around the dimer solely by moving the atom in the bottom left corner further away from the bright defect (Suppl. Fig.\,\ref{fig:switching}b). This was precisely the atom that we replaced in the first step to induce the switch from CC to HC. It, thus, clearly reveals the reversibility of the switch of the CDW domain structure. Repositioning the other atoms does not change the CDW-to-lattice alignment close to the dimer as shown in Suppl. Fig.\,\ref{fig:switching}c. We therefore assume, than one could in principle switch between both CDW to lattice alignments solely by manipulating one additional Fe atom if the other pinning Fe atoms and defects are positioned favorably.

\begin{figure}\centering	\includegraphics[width=\linewidth]{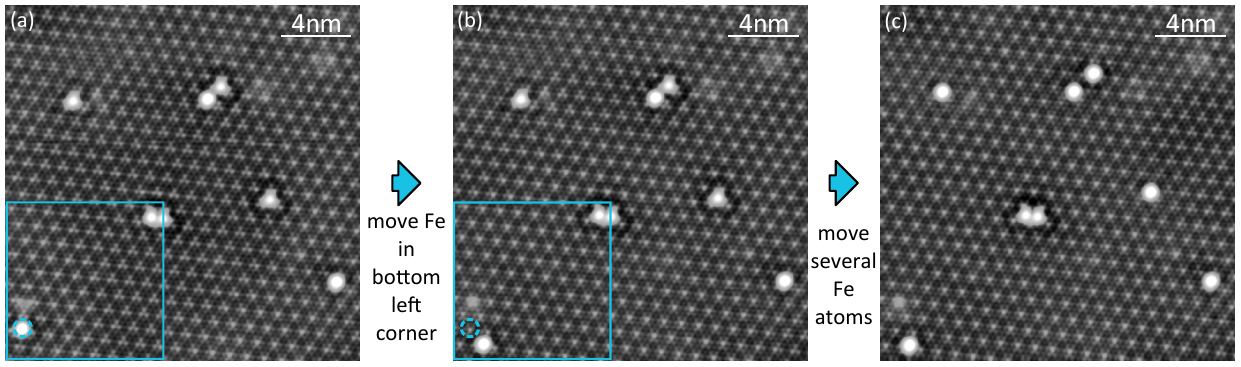}
	\caption{Manipulation of CDW around 2$a$ dimer: Three stages of manipulating the CDW around the $2a$ dimer from HC to CC alignment. $\Delta_{\mathrm{tip}}\approx$1.44\,mV; set point: 10\,mV, 50\,pA.
	}
	\label{fig:switching}
\end{figure}

\begin{figure}\centering	\includegraphics[width=\linewidth]{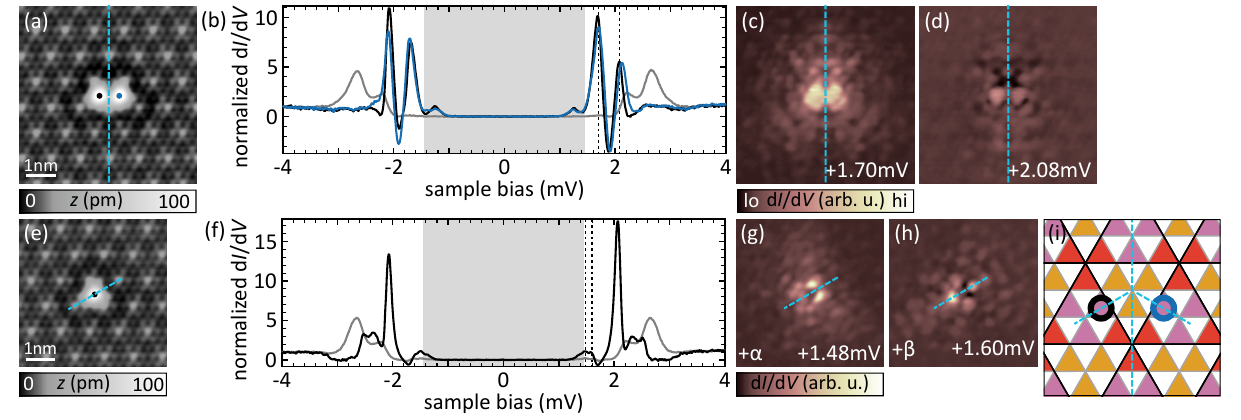}
	\caption{2$a$ dimer and monomer in CC domain: (a, e) Topographic images of the $2a$ dimer and the monomer after switching the CDW back to CC. (b, f) Spectra recorded on both dimer atoms and the monomer. (c, d, g, h) \didv maps of the two resonances lowest in energy of the $2a$ dimer (c, d) and the monomer (g, h) after the CDW was switched back. (i) Schematic of the \nbse\,surface in an CC CDW area. Hollow sites are color coded by their position with respect to the CDW (black lines) and the adsorption sites of the dimer atoms are indicated by color coded circles.
	$\Delta_{\mathrm{tip}}\approx$1.44\,mV; Set point: (a),(e) 10\,mV, 100\,pA; rest 5\,mV, 750\,pA; all: V$_\mathrm{rms}$=15\,$\mu$V.
	}
	\label{fig:CCafter}
\end{figure}

We once again record a dataset on the $2a$ dimer after switching back the CDW to the CC domain and show it in Suppl. Fig.\,\ref{fig:CCafter}a-d. We then remove the atom marked by the blue dot and reproduce the monomer shown in Fig.\,2 of the main text (Suppl. Fig.\,\ref{fig:CCafter}e-h). Both datasets perfectly reproduce the ones shown in Suppl. Fig.\,\ref{fig:CCbefore} for the $2a$ dimer and Fig.\,2a-d of the main text for the monomer proving that the changes caused by manipulating the CDW are indeed fully reversible.

\section{Supplementary Note 7: Detailed data on the switching processes}
\begin{figure}\centering	\includegraphics[width=\linewidth]{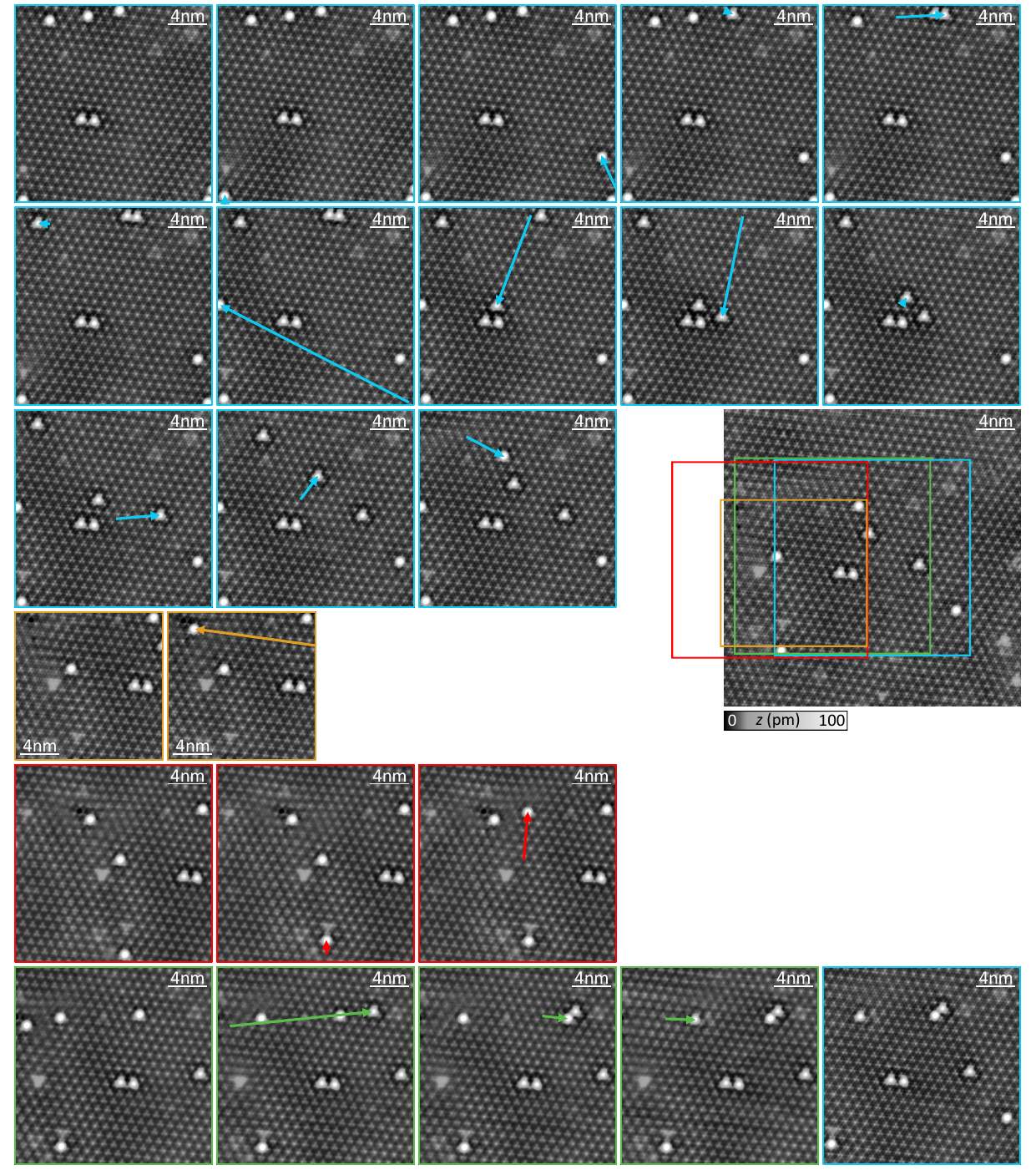}
	\caption{Topographic images of each manipulation step when manipulating the CDW from CC to HC (order left to right, top to bottom). Different scan frames are indicated by color-coded frames.
		Set point: 10\,mV, 50\,pA.
	}
	\label{fig:switching1all}
\end{figure}

Here, we show the topographic images recorded after each manipulation step for both CDW manipulations shown in the main text. We indicate which atom was moved by arrows. Supplementary Fig.\,\ref{fig:switching1all} shows the manipulation of the CDW from CC to HC alignment around the $4a$ dimer. The movement of each atom is indicated by arrows in the image after manipulation of an Fe atom. The sequence of images shows that in some cases the manipulation of Fe atoms does not lead to any changes while in other cases a small displacement of a single atom leads to a switch. The complexity of the processes reveals the delicate balance of the local minima in the energy landscape of the CDW domains.

\begin{figure}\centering	\includegraphics[width=\linewidth]{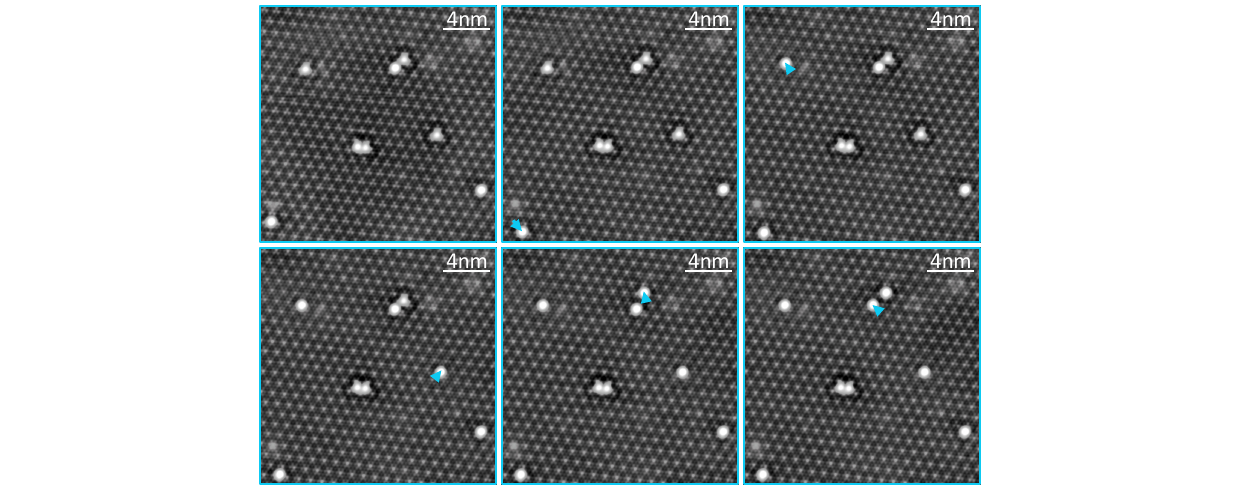}
	\caption{Topographic images of each manipulation step when manipulating the CDW back to CC alignment from left to right and top to bottom.
		Set point: 10\,mV, 50\,pA.
	}
	\label{fig:switching2all}
\end{figure}

When we manipulated the CDW back to the CC alignment (all images in Suppl. Fig.\,\ref{fig:switching2all}), we needed a significantly smaller number of steps due to the understanding of the influence of the individual Fe atoms.

\end{document}